\documentclass{ws-procs9x6}

\begin{document}

\title{Microscopic Description of Nuclear Fission:\\
Fission Barrier Heights of Even-Even Actinides}

\author{J. McDonnell$^*$ and N. Schunck}

\address{Physics Division, Lawrence Livermore National Laboratory, \\
7000 East Ave, Livermore, CA 94550/USA\\
%E-mail: an\_author@laboratory.com
$^*$E-mail: mcdonnell5@llnl.gov}

\author{W. Nazarewicz}
\address{Department of Physics \&
Astronomy, University of Tennessee, \\Knoxville, Tennessee 37996, USA\\
Physics Division, Oak Ridge National Laboratory, \\Oak Ridge, Tennessee 37831, USA\\
Institute of Theoretical Physics, University of Warsaw,  00-681 Warsaw, Poland
}

\begin{abstract}
We evaluate the performance of modern nuclear energy density functionals for predicting inner and outer fission barrier heights and  energies of fission isomers of even-even actinides. 
%We investigate whether the zero-point correction helps to improve agreement between theoretical and empirical inner barrier heights. 
For isomer energies and outer barrier heights, we find that the
self-consistent theory at the HFB level
is capable of providing quantitative agreement  with  empirical data.  
\end{abstract}


\bodymatter

\section{Introduction}\label{section:intro}

Attaining the goal of a fully microscopic theory of fission has been greatly helped by advances in the nuclear density functional theory (DFT) and recent developments in the field of high-performance computing. A reliable microscopic theory of fission based solely on the properties of the underlying effective nuclear interactions would offer unprecedented predictive power, thereby giving us more  confidence in tackling important questions related, e.g., to properties of  superheavy elements and astrophysical $r$-process.  

In the recent years,there have been systematic studies of nuclear fission in the DFT framework. In particular, the zero-range Skyrme functional \cite{ErlerPRC85,StaszczakSHE}, the finite-range Gogny force \cite{PhysRevC.66.014310,WardaSHE} and the relativistic mean-field approaches \cite{AbusaraPRC82,LuPRC85} have been used to compute barrier and half-live estimates. In parallel, the optimization of coupling constants in energy density functionals \cite{KortelPRC82} has been revisited with a particular focus on uncertainty quantification and correlation analysis. In this context, the \textsc{UNEDF1} parametrization of the Skyrme functional \cite{KortelainenPRC85} was recently developed with fission applications in mind, optimized to have the realistic deformation properties \cite{NikolovPRC83}.

The goal of this work is to assess the predictive power of the UNEDF1 functional in fission studies by comparing its predictions to experimental data such as barrier heights and isomer excitation energies, and benchmarking it against alternative models. We briefly outline the nuclear DFT formalism for fission in Sec.~\ref{section:theory}. We then present and discuss our results in Sec.~\ref{section:barriers}, before concluding in Sec.~\ref{section:discussion}.  

\section{Theory and Background}\label{section:theory}

We calculate potential energy surfaces through nuclear DFT  with constraints, at the level of the Hartree-Fock-Bogoliubov (HFB) mean field. In the particle-hole channel, we use the \textsc{UNEDF1} functional and the SkM$^\ast$ functional as a benchmark \cite{BartelNPA386}. In the particle-particle channel, we use the density-dependent mixed-pairing interaction \cite{Dobaczewski02}. All calculations were performed with a quasi-particle cutoff energy of $E_{\rm cut} = 60$\,MeV. The \textsc{UNEDF1} functional prescribes the pairing strengths; for our calculations with SkM$^\ast$, we determined them such that the calculated pairing gaps coincide with the experimental odd-even mass differences in $^{232}$Th. 

We explore configurations that explicitly break axial symmetry, to properly account for the inner barrier height, as well as reflection symmetry, to minimize the outer barrier height. Calculations are performed with the symmetry-unrestricted nuclear DFT solver HFODD \cite{SchunckCPC2012}, which uses a harmonic oscillator (HO) basis. We use the lowest 1140 states from 31 deformed HO shells.  For each value of the quadrupole moment $Q_{20}$, the basis is deformed according to $\alpha_{20} = 0.05 \sqrt{Q_{20}}$, and the basis frequency is set to $\hbar \omega_0 = 6.5 + 0.1 \times Q_{20} e^{-0.02 Q_{20}}$. This empirical parametrization gives a good compromise between accuracy and speed.\cite{SchunckZak2012}

\section{Fission Barrier Heights}\label{section:barriers}

While fission barrier heights are not observables  in the strict sense, they do provide important constraints on theories of fission. Our goal is to evaluate and compare the ability of theoretical models to predict quantities such as fission barrier heights beyond known data. 

\subsection{Comparison of theoretical models}\label{subsection:comparingUNEDF1}

\begin{figure}[!h]
  \centering
  \includegraphics[width=0.65\textwidth]{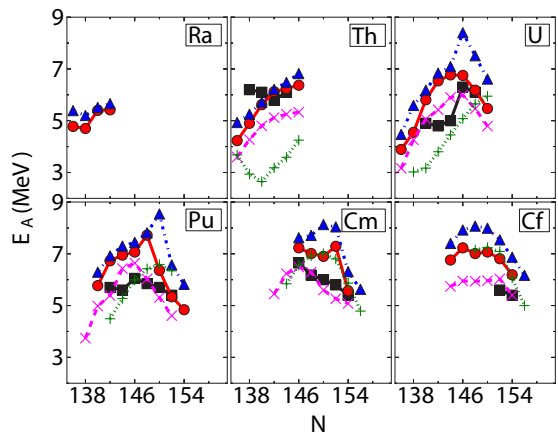}
  \caption[A comparison between empirical values and predicted values for the inner barrier heights of several actinides]{Inner barrier heights of selected even-even  actinide nuclei. Experimental data \cite{CapoteNDS110,SmirenkinVienna1993}
  (black lines with square markers) are compared to predictions of  \textsc{UNEDF1} (red lines with circles);  SkM$^\ast$  (blue lines with triangles); D1S  \cite{Delaroche2006103} (magenta lines with `x'-es); and FRLDM 
\cite{MollerPRC79} (green lines with plus signs).}  
  \label{fig:InnerByChain}
 \end{figure}

\begin{figure}[!h]
  \centering
  \includegraphics[width=0.65\textwidth]{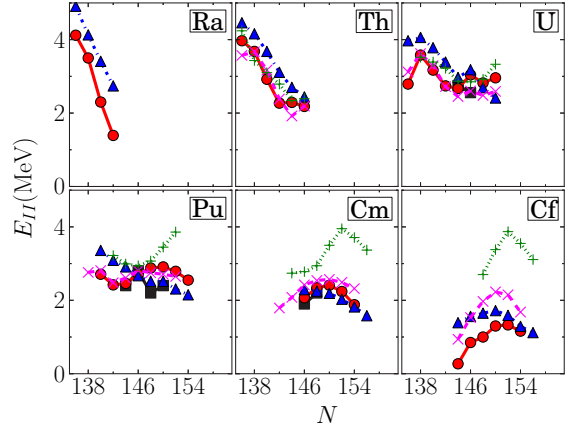}
  % hg180T10contour.eps: 0x0 pixel, 300dpi, 0.00x0.00 cm, bb=
  \caption[A comparison between empirical values and predicted values for the fission isomer energies of several actinides]{Similar to Fig.~\ref{fig:InnerByChain} but for bandhead energies  of
 fission isomers.  The experimental data are from  Ref.~\cite{SinghNDS2002}.  }  
  \label{fig:IsomerByChain}
 \end{figure}

\begin{figure}[!h]
  \centering
  \includegraphics[width=0.65\textwidth]{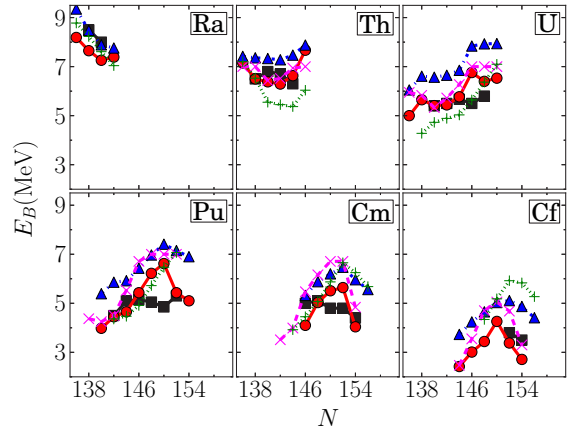}
  % hg180T10contour.eps: 0x0 pixel, 300dpi, 0.00x0.00 cm, bb=
  \caption[A comparison between empirical values and predicted values for the outer barrier heights of several actinides]{Similar to Fig.~\ref{fig:InnerByChain} but for the  outer barrier heights.}  
  \label{fig:OuterByChain}
 \end{figure}

In a first step, we benchmark the UNEDF1 functional against alternative parametrizations of the energy functional (Skyrme SkM$^\ast$ and Gogny D1S \cite{Delaroche2006103}) and the microscopic-macroscopic FRLDM \cite{MollerPRC79} models for the inner fission barrier height (Fig.~\ref{fig:InnerByChain}), the energy of the superdeformed fission isomer (Fig.~\ref{fig:IsomerByChain}), and the outer fission  barrier height (Fig.~\ref{fig:OuterByChain}).  

For the inner barrier height $E_{A}$, we note that the three microscopic approaches exhibit the same isotopic trend, with a roughly parabolic  $E_{A}(N)$ behavior, and differ essentially by the position of the maximum in the U, Pu and Cm isotopes. While FRLDM results are similar to the DFT results in the heavy actinides, they show marked differences in the lighter systems, especially Th and U.  Overall, the excitation energy of the fission isomer is much more consistent across all theoretical models used here, with the exception of the Cm and Cf isotopes  where the FRLDM values are as much as twice the DFT value (and the empirical data). Most interestingly, we note that all the models considered agree better at predicting $E_{B}$ than  $E_{A}$: the RMS of the variances about the mean theoretical prediction for each element ranges between 1.32 MeV (Cm) to 1.88 MeV (Cf) for $E_{B}$, while it ranges from 1.65 MeV (Cm) to 2.74 MeV (U) for $E_{A}$. This is somewhat counterintuitive, as one would think that the difference between the deformation properties of these approaches would be magnified as the elongation increases. 

\subsection{Comparison with empirical values}\label{subsection:compareEmpirical}

\begin{figure}[!h]
  \centering
  \includegraphics[width=0.65\textwidth]{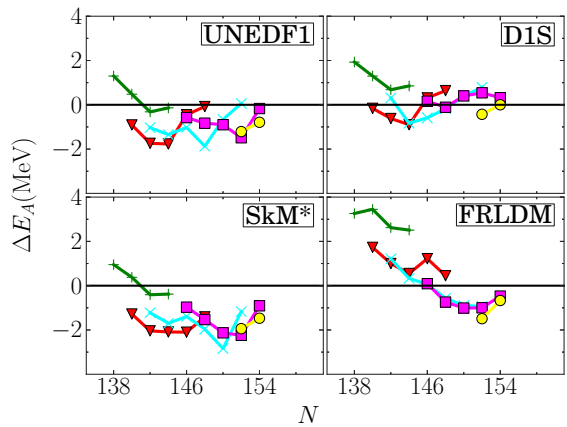}
  \caption[The residuals of the theoretical predictions for the inner barrier heights]{The residuals $\Delta E $ for the inner barrier heights  of selected even-even  Th (green lines with plus signs), U (red lines with inverted trianges), Pu (cyan lines with `x'-es), Cm  (magenta lines with squares), and Cf (yellow lines with circles) isotopes.}  
  \label{fig:Inner_resid}
 \end{figure}

\begin{figure}[!h]
  \centering
  \includegraphics[width=0.65\textwidth]{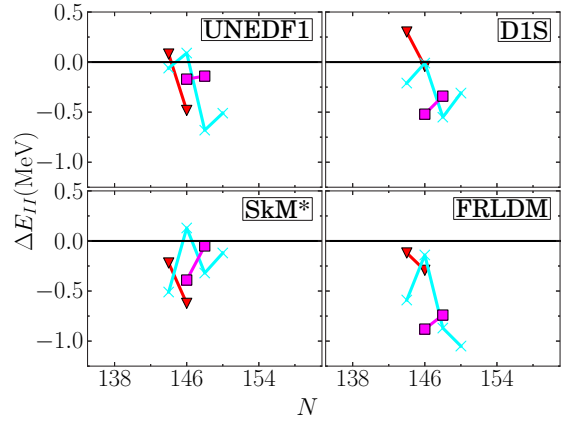}
  % hg180T10contour.eps: 0x0 pixel, 300dpi, 0.00x0.00 cm, bb=
  \caption[The residuals for the fission isomer energies]{Similar to Fig.~\ref{fig:Inner_resid} but for bandhead energies  of
 fission isomers.}  
  \label{fig:Isomer_resid}
 \end{figure}

\begin{figure}[!h]
  \centering
  \includegraphics[width=0.65\textwidth]{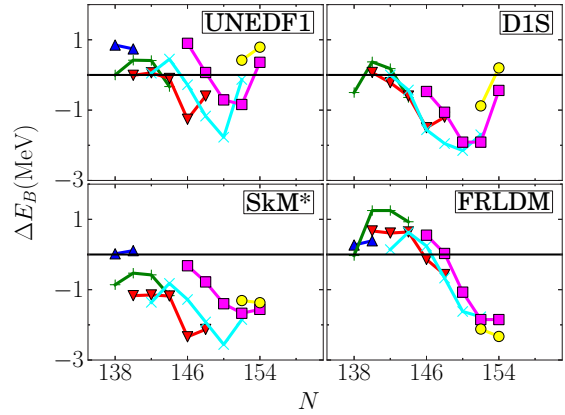}
  % hg180T10contour.eps: 0x0 pixel, 300dpi, 0.00x0.00 cm, bb=
  \caption[The residuals for the outer barrier heights]{Similar to Fig.~\ref{fig:Inner_resid} but for outer barrier heights.  The additional blue lines with triangles mark  Ra isotopes.}  
  \label{fig:Outer_resid}
 \end{figure}

We now turn to the residuals $\Delta E = E_{\text{exp}} - E_{\text{th}}$, with the empirical values taken from Refs.~\cite{CapoteNDS110,SmirenkinVienna1993}.
When assessing these results one needs to bear in mind that the empirical uncertainty on the barrier heights is at least 0.3\,MeV.
The residuals for the inner barrier heights are presented in Fig.~\ref{fig:Inner_resid} for \textsc{UNEDF1}, SkM$^\ast$, D1S and FRLDM. Both  \textsc{UNEDF1} and SkM$^\ast$   systematically overestimate $E_{A}$  and exhibit a gentle oscillating trend. The FRLDM residuals have a downward overall slope with $N$, which is especially pronounced for the U and Pu chains. The D1S residuals have the least scatter for inner barrier heights. The residuals for  the fission isomer excitation  energy are shown in Fig.~\ref{fig:Isomer_resid}. There are but a few measured data points, so we simply note that the self-consistent models exhibit less scatter than FRLDM. Finally, we show the residuals for outer barrier heights in Fig.~\ref{fig:Outer_resid}. The \textsc{UNEDF1} functional demonstrates the least scatter, while both D1S and SkM$^\ast$ consistently overestimate the outer barrier heights. In addition, the FRLDM residuals exhibit  a downward slope with $N$, which is similar to the trend for $E_{A}$.

In this comparison with empirical fission barrier heights and fission isomer energies, the performance of \textsc{UNEDF1} is on par with that of the other self-consistent interactions and the macroscopic-microscopic FRLDM. For  the four theoretical models discussed in this work,  Table~\ref{tab:Actinide_RMS} displays the RMS deviations from experiment for the even-even isotopes.
For inner barrier heights, the D1S functional yields the best agreement, with an RMS of $0.709$ MeV, but \textsc{UNEDF1} agrees nearly as well with an RMS of $1.03$ MeV, i.e., these two models are practically equivalent considering
the empirical uncertainty on barrier heights. The three self-consistent calculations yield a similar level of agreement for the fission isomer energy, while for the outer barrier heights \textsc{UNEDF1} is the closest to the empirical values, with an RMS of $0.690$ MeV. The FRLDM and D1S predictions for outer barrier heights are comparable.

\begin{table}[!h]
\tbl{RMS deviations of the first barrier height $E_A$, the fission isomer $E_{II}$, and the second barrier height $E_B$ are given for the selection of even-even nuclei considered.}
{
\begin{tabular}{@{}lcccc@{}}
\toprule
× & \textsc{UNEDF1} & FRLDM & SkM$^\ast$ & D1S\\ \colrule
$E_A$ [MeV] & 1.03 & 1.52 & 1.61 & 0.709\\ %\hline
$E_{II}$ [MeV] & 0.357 & 0.675 & 0.351 & 0.339\\ %\hline
$E_B$ [MeV] & 0.690 & 1.13 & 1.39 & 1.14\\ \botrule
\end{tabular}
}
\label{tab:Actinide_RMS}
\end{table}

It is to be noted that the inner barrier heights calculated with {\it all} microscopic models are consistently $0.5$--$1.0$ MeV too high for U and Pu isotopes, and are often higher than typical values of one-neutron separation energy (about $6.5$ MeV). Using such numbers directly in reaction models to compute the fission spectrum, would lead to, e.g.,  $^{240}$Pu being non-fissile. We should bear in mind, however, the large uncertainty affecting the empirical data, the simplifying assumptions used in reaction models, and the systematic and statistical errors of DFT calculations. On the DFT side, basis truncation errors \cite{SchunckZak2012} and beyond mean-field corrections can modify the potential energy surface in a non trivial way \cite{StaszczakSHE,Delaroche2006103}. If the zero-point energy is taken into account according to the method of Ref.~\cite{StaszczakSHE}, for example, the inner fission barrier heights can be lowered by approximately $0.5$\,MeV.  Furthermore, neutron capture leads to  compound nuclei in highly excited states.
The excitation energy $E^*$ has the effect of lowering the fission barriers as well \cite{SheikhPRC80} so one needs to be careful when using the ground-state barriers ($E^*=0$) in estimates of (n,f) cross sections.

\section{Conclusions}\label{section:discussion}

In this survey of  fission barrier properties of even-even actinides, we find that the recently developed  \textsc{UNEDF1} energy density functional yields predictions that agree well with experimental values and are on a par with, or better, than predictions of other self-consistent or macroscopic-microscopic models. While the inner barrier heights are systematically  overestimated, one also needs to bear in mind that empirical values are  subject to error that is at least 0.3\,MeV. As fit-observables  are selected for future optimizations, it will be valuable to consider if additional data such as one-neutron separation energy may help to better constrain fission barrier heights.

\section*{Acknowledgements}

Useful discussions with A. Baran, A. Staszczak, and P. Talou are gratefully acknowledged. This work was supported by the U.S. Department of Energy under Contract Nos.\ DE-AC52-07NA27344 (Lawrence Livermore National Laboratory),  DE-FG02-96ER40963 (University of Tennessee), DE-FG52-09NA29461 (the Stewardship Science Academic Alliances program), DE-AC07-05ID14517 (NEUP grant sub award 00091100), and DE-SC0008499 (NUCLEI SciDAC Collaboration). An award of computer time was provided by the Innovative and Novel Computational Impact on Theory and Experiment (INCITE) program. This research used resources of the Oak Ridge Leadership Computing Facility located in the Oak Ridge National Laboratory, which is supported by the Office of Science of the Department of Energy under Contract DE-AC05-00OR22725.

\bibliographystyle{ws-procs9x6}

\begin{thebibliography}{10}

\bibitem{ErlerPRC85}
J.~Erler, K.~Langanke, H.~P. Loens, G.~Mart{\'\i}nez-Pinedo and P.-G. Reinhard,
  {\em Phys. Rev. C} {\bf 85}, p. 025802 (2012).
  
\bibitem{StaszczakSHE}  
A. Staszczak, A. Baran, and W. Nazarewicz, {\em arXiv/1208.1215} (2012).

\bibitem{PhysRevC.66.014310}
M.~Warda, J.~L. Egido, L.~M. Robledo and K.~Pomorski, {\em Phys. Rev. C} {\bf
  66}, p. 014310 (2002).

\bibitem{WardaSHE}  
M.~Warda and  J.~L. Egido, {\em Phys. Rev. C} {\bf 86}, p. 014322 (2012). 

\bibitem{AbusaraPRC82}
H.~Abusara, A.~V. Afanasjev and P.~Ring, {\em Phys. Rev. C} {\bf 82}, p.
  044303 (2010).

\bibitem{LuPRC85}
B.-N. Lu, E.-G. Zhao and S.-G. Zhou, {\em Phys. Rev. C} {\bf 85}, p. 011301 (2012).

\bibitem{KortelPRC82}
M.~Kortelainen, T.~Lesinski, J.~Mor{\'e}, W.~Nazarewicz, J.~Sarich, N.~Schunck,
  M.~V. Stoitsov and S.~Wild, {\em Phys. Rev. C} {\bf 82}, p. 024313 (2010).

\bibitem{KortelainenPRC85}
M.~Kortelainen, J.~McDonnell, W.~Nazarewicz, P.-G. Reinhard, J.~Sarich,
  N.~Schunck, M.~V. Stoitsov and S.~M. Wild, {\em Phys. Rev. C} {\bf 85}, p.
  024304 (2012).

\bibitem{NikolovPRC83}
N.~Nikolov, N.~Schunck, W.~Nazarewicz, M.~Bender and J.~Pei, {\em Phys. Rev. C}
  {\bf 83}, p. 034305 (2011).

\bibitem{BartelNPA386}
J.~Bartel, P.~Quentin, M.~Brack, C.~Guet and H.-B. H{\aa}kansson, {\em Nucl.
  Phys. A} {\bf 386}, p. 79 (1982).
  
\bibitem{Dobaczewski02}
J.~Dobaczewski, W.~Nazarewicz and M.~V. Stoitsov, {\em Eur. Phys. J. A} {\bf
  15}, p. 21 (2002).

\bibitem{SchunckCPC2012}
N.~Schunck, J.~Dobaczewski, J.~McDonnell, W.~Satu{\l}a, J.~Sheikh,
  A.~Staszczak, M.~Stoitsov and P.~Toivanen, {\em Comput. Phys.
  Commun.} {\bf 183}, p. 166 (2012).

\bibitem{SchunckZak2012}
N.~Schunck, {\em arXiv/1212.3356} (2012).

\bibitem{Delaroche2006103}
J.-P. Delaroche, M.~Girod, H.~Goutte and J.~Libert, {\em Nucl. Phys. A}
  {\bf 771}, p. 103 (2006).

\bibitem{MollerPRC79}
P.~M{\"o}ller {\em et~al.}, {\em Phys. Rev. C} {\bf 79}, p. 064304 (2009).

\bibitem{CapoteNDS110}
R.~Capote {\em et~al.}, {\em Nucl. Data Sheets} {\bf 110}, p. 3107 (2009).

\bibitem{SmirenkinVienna1993}
G.~Smirenkin, {\em INDC(CCP)-359}, Tech. Rep., IAEA  (1993).

\bibitem{SinghNDS2002}
B.~Singh, R.~Zywina and R.~B. Firestone, {\em Nucl. Data Sheets} {\bf 97},
  p. 241 (2002).

\bibitem{SheikhPRC80}
J.~A.~Sheikh, W.~Nazarewicz, and J.C.~Pei, {\em Phys. Rev. C} {\bf 80}, p. 011302 (2009).

\end{thebibliography}

\end{document}